Kari D. Weaver

# The Artificial Intelligence Disclosure (AID) Framework

## An Introduction

As artificial intelligence (AI) tools—particularly generative AI based in large language models—are becoming widely available, their use across the varied contexts of education, work, and research must be negotiated. The accelerating uptake of these tools is driving a range of conversations around transparency in the use of these tools for various purposes.[1]

Within the contexts of education and research, and particularly within higher education, the citation has long been the standard tool for providing transparency and connection in the transfer of ideas across scholars, framing of arguments, and design of methodologies.[2] Accordingly, as AI tools have grown in prominence, organizations that publish style manuals and guides have provided citation guidance to address the use of AI-generated content to inform education and research practice.[3] The ACRL Framework for Information Literacy for Higher Education reinforces this practice through the themes Information Creation as a Process and Scholarship as Conversation, which directly address citation practices as an element of information literacy.[4]

Unfortunately, citations do not fully meet the needs of today's AI-enabled world. Citations emphasize the fixed form of a tangible output.[5] This is incongruent with today's generative AI systems, where the specific interplay of prompt, model, and model parameters creates a unique output that is not always repeatable, reproducible, or recallable, depending on the technology. Citations also focus on the ideas posed by an author, whereas generative AI can serve a variety of meaningful functions in the writing process, including researcher, editor, critic, collaborator, and more. While today's citation practices do help provide some transparency, they are not sufficient, in and of themselves, to capture the varied ways AI tools function or are being used across contexts.

In response, there have been burgeoning conversations and recommendations around the need to attribute the use of AI in research.[6] Thus far, this need has been met via the recommended inclusion of a note, with little to no guidance on what the note itself should include. This has been identified as a problem to the use of AI in academic and research contexts. Calls for greater transparency and granularity in the use of AI abound.[7]

Despite the growing calls for increased transparency,[8] a gap currently exists across all contexts regarding how and to what level of detail disclosure of AI tools require.[9] While this is a particular issue within the research community, where the authors are still to be held firmly to account for their work, it relates back to academic integrity concerns within education and has implications for student assessment and instructional practice.

Kari D. Weaver is learning, teaching, and instructional design librarian at the University of Waterloo, email: kdweaver@uwaterloo.ca.

In the service of addressing this gap, across all those varied contexts, I would like to introduce the Artificial Intelligence Disclosure (AID) Framework. The AID Framework was inspired by the Contributor Roles Taxonomy (CRediT) developed in 2012 through a collaborative workshop hosted by the Wellcome Trust and Harvard University.[10] The purpose of a CRediT statement is to outline contributors' individual roles as authors within a research output.[11] The AID Framework adapts this concept to the use of AI.

My approach to the AID framework is grounded in my own professional work as an academic librarian, my experiences as a scholar and educator, and my engagement as a member of the University of Waterloo's Associate Vice-President Academic's Standing Committee on New Technologies, Pedagogy, and Academic Integrity. It is meant to provide transparency to the use of AI tools throughout the writing process, ensuring clarity at a level that is both detailed enough to be informative and short enough to avoid being onerous. While the specific taxonomy described below is targeted toward academic and research use, it can easily be adapted to other contexts and workflows where disclosure of AI use is important.

## Artificial Intelligence Disclosure (AID) Framework

The purpose of the Artificial Intelligence Disclosure (AID) Framework is to provide brief, targeted disclosure about the use of AI systems based on the range of activities used for research writing. The AID Statement is appended to the end of the paper (similar to an acknowledgments section), detailing the AI tools used and the manner in which they were used, based on the possible points of engagement through the writing process, as captured in the headings below. As generative AI tools may not be an author of scholarly work, overlap in categorization between CRediT and AID Framework have been edited as necessary to reflect this distinction. The formatting is intended to be both human- and machine-readable, and uses the following structure:

> **AID Statement**: *Artificial Intelligence Tool*: [description of tools used]; *[Heading]:* [description of AI use in that stage of the work]:

Each heading: statement pair will end in a semi-colon, except for the last statement, which will end in a period. Any other symbols can be used in the "statement" portion of the heading: statement pair except for colons and semi-colons.

If AI tools were used at any point in the writing, research, or project management processes, the AID Statement will always begin with the "artificial intelligence tool" section. It will then be followed by any heading: statement pairs necessary to disclose AI tool use. Heading: statement pairs will only be included if AI was used in that portion of the writing process. If a heading is not needed, it should not be included. If AI was not used at any point in the writing, research, or project management processes, authors would not include an AID Statement in their work.

The potential headings for the AID Statement, and their definitions, are the following:

1. *Artificial Intelligence Tool(s)*: The selection of tool or tools and versions of those tools used and dates of use. May also include note of any known biases or limitations of the models or data sets.
2. *Conceptualization*: The development of the research idea or hypothesis including framing or revision of research questions and hypotheses.

3. *Methodology*: The planning for the execution of the study including all direct contributions to the study design.

4. *Information Collection*: The use of AI to surface patterns in existing literature and identify information relevant to the framing, development, or design of the study.

5. *Data Collection Method*: The development or design of software or instruments used in the study.

6. *Execution*: The direct conduct of research procedures or tasks (e.g. AI web scraping, synthetic surveys, etc.)

7. *Data Curation*: The management and organization of those data.

8. *Data Analysis*: The performance of statistical or mathematical analysis, regressions, text analysis, and more using AI tools.

9. *Privacy and Security*: The ways in which data privacy and security were upheld in alignment with the expectations of ethical conduct of research, disciplinary guidelines, and institutional policies.

10. *Interpretation*: The use of AI tools to categorize, summarize, or manipulate data and suggest associated conclusions.

11. *Visualization*: The creation of visualizations or other graphical representations of the data.

12. *Writing—Review & Editing*: The revision and editing of the manuscript.

13. *Writing—Translation*: The use of AI to translate text across languages at any point in the drafting process.

14. *Project Administration*: Any administrative tasks related to the study, including managing budgets, timelines, and communications.

The following are examples of AID Statements and their usage for research and education.

**For Use in Research**

Researchers need detailed guidance to fully articulate the variety and depth of ways AI tools have been used throughout the research and publication processes. The AID Framework, as exemplified in the following sample AID Statement, can address this need in a clear and focused manner.

*AID Statement Example*

*Artificial Intelligence Tool*: ChatGPT v.4o and Microsoft Copilot (University of Waterloo institutional instance); *Conceptualization*: ChatGPT was used to revise research questions; *Data Collection Methods:* ChatGPT was used to create the first draft of the survey instrument; *Data Analysis*: Microsoft Copilot was used to verify identified themes coded from open ended survey responses; *Privacy and Security*: no identifiable data was shared with ChatGPT during the design of this study, only the University of Waterloo institutional instance of Microsoft Copilot was used to analyze any anonymized research data in compliance with University of Waterloo privacy and security policies; *Writing—Review & Editing*: ChatGPT was used in the literature review to provide sentence-level revisions and metaphor options; *Project Administration*: ChatGPT was used to establish a list of tasks and timelines for the study.

**For Use in Education**

Within educational settings, the AID Framework assists with openly articulating the use of AI tools in student work, although a less extensive and detailed disclosure is appropriate. For instance, the following example could be used in a kinesiology student research paper examining the effectiveness of motor-performance fitness tasks on sedentary office worker health.

*AID Statement Example*

*Artificial Intelligence Tool*: Microsoft Copilot (University of Waterloo institutional instance); *Conceptualization*: Microsoft Copilot was used to identify key motor-performance fitness tasks in the development of the research question; *Information Collection*: I used Microsoft Copilot to find relevant journal articles and other sources; *Visualization*: I used Microsoft Copilot to create a graph comparing the different motor-performance fitness tasks included in my paper; *Writing—Review & Editing:* I used Microsoft Copilot to help break down my paragraph-long draft sentences into clearer, shorter ones.

Beyond use in student work, the AID Framework is also helpful for transparent disclosure of the use of AI tools for instructional tasks that are increasingly automated including lesson planning, rubric creation, and curriculum mapping.[12] Consequently, instructors may wish to incorporate an AID Statement directly within their instruction or assessment materials and can adapt an AID Statement for use in a course syllabus or learning management system.

## Conclusion

The AID Framework provides a method for transparency of AI use in writing that is clear, consistent, succinct, and amenable to both human and machine use. It can also be adapted to a range of other contexts where they are consistent and enable AI use at multiple points in a workflow. While AI is a fast-developing field with growing capabilities, this structured approach allows us to do our best work faster and more efficiently without losing sight of the critical human additions. Finally, adopting a consistent approach to AI disclosure through the AID Framework simplifies the expectations and needed elements to maintain academic and research ethics.

## Acknowledgments

Working to develop and articulate the varied aspects and use cases of AI to inform the AID Framework has not been a solitary task. Though it is my own, I have been supported in the development of the AID Framework by my partner, Dulany Weaver, an expert in AI. Colleagues at the University of Waterloo, particularly Nadine Fladd, Karen Lochead, Amanda McKenzie, and Trevor Holmes, also provided thoughtful and encouraging feedback during the development process.

## Notes

1. "Transparent Use of Artificial Intelligence," Martine Peters, chercheuse et professeure en science de l'éducation Directrice du Partenariat universitaire sur la prévention du plagiat, accessed June 15, 2024, https://mpeters.uqo.ca/logos-ai-en-peters-2023/.